\documentclass[preprint]{vgtc}
\ifdefined\pdfminorversion\pdfminorversion=7\fi
\usepackage[T1]{fontenc}
\usepackage[utf8]{inputenc}
\usepackage{newtxtext,newtxmath}
\usepackage[shrink=25]{microtype}
\usepackage{amsmath}
\usepackage{cite}
\usepackage{xurl}
\usepackage{hyperref}
\usepackage{stfloats}
\usepackage{flushend}
\preprinttext{Preprint}
\nocopyrightspace
\graphicspath{{figures/}}
\hypersetup{
  hidelinks,
  pdftitle={WL-Headset Law: Projecting Future Headset Evolution from a Decade of Display and Mass Data},
  pdfauthor={Jialin Wang and Hai-Ning Liang},
  pdfsubject={Angular detail and head-worn mass in head-mounted displays},
  pdfkeywords={Head-mounted displays, virtual reality, augmented reality, pixels per degree, head-worn mass}
}
\title{WL-Headset Law: Projecting Future Headset Evolution from a Decade of Display and Mass Data}
\author{Jialin Wang$^{1}$ \and Hai-Ning Liang$^{1,2}$\thanks{Corresponding author. E-mail: \href{mailto:hainingliang@hkust-gz.edu.cn}{\texttt{hainingliang@hkust-gz.edu.cn}}}}
\affiliation{\scriptsize
$^{1}$Computational Media and Arts Thrust, The Hong Kong University of Science and Technology (Guangzhou), Guangzhou, China\\
$^{2}$Division of Arts and Machine Creativity, The Hong Kong University of Science and Technology, HKSAR, China}
\abstract{Tracking a decade of headset display and mass changes can inform forecasts. However, comparing headset resolution requires pixels per degree (PPD), which is often unreported and estimated using unclear assumptions. Therefore, we compile the open-source, evidence-linked HMD Resolution and Mass Database (HRMD) with 241 configurations from 2016--2026 and introduce structure- and axis-aware robust calibration (SARC) to estimate missing central PPD. Compared with official PPD, SARC reduces PPD estimation error from 14.8\% for traditional geometric estimation to 3.2\%. Based on HRMD, we introduce the Weight-Linked Headset Law (WL-Headset Law), which models PPD by year, mass, and category. The law reveals 7.4\% annual PPD growth at fixed mass and category. Meanwhile, head-worn mass fell by 6.5\% annually after category adjustment. At an eyeglass-like mass of 30~g, the law projects average central PPD of 60 (normal visual acuity) around 2035 and 120 (higher acuity) around 2044 if trends persist.}
\begin{document}
\maketitle

\section{Introduction}\label{sec:introduction}
A central goal for augmented reality (AR) and virtual reality (VR) headsets is to improve visual resolution while reducing head-worn mass. In VR, resolution affects visual experience and can influence simulator sickness \cite{Wang2022RenderResolution,Saredakis2020VRSickness}. Head-worn mass, mass distribution, and task posture affect the wearer's physical load \cite{Chihara2018PhysicalWorkload,Ito2021WeightBalance}. Advances in displays and near-eye optics offer routes to higher resolution and more compact hardware \cite{Xiong2021Displays,Hsiang2022LightEngines,Koulieris2019DisplayTracking}. These advances raise a practical question: how soon could headsets combine high visual resolution with an eyeglass-like mass? To estimate this timescale, we analyze a decade of headset specifications and extrapolate the relationship between resolution, head-worn mass, and year, assuming historical trends persist.

Comparing resolution across headsets requires accounting for the field of view (FoV) over which their pixels are distributed. Pixels per degree (PPD) measures the number of display pixels per degree of visual angle. For a fixed pixel count, a wider FoV means a lower average PPD, while optical design affects how PPD varies from the center to the edges. A headset with more pixels can therefore provide a wider view without a proportional increase in central detail.

Making this comparison across devices requires resolving gaps in published specifications. Only 23.0\% of the records in our open-source HMD Resolution and Mass Database (HRMD) have official PPD values, and many reported FoVs omit their axis. Traditional geometric estimation for PPD divides pixels along an axis by the corresponding FoV, giving an average angular density that can differ from central PPD. We therefore introduce structure- and axis-aware robust calibration (SARC), which estimates missing central PPD from display and FoV specifications using corrections learned from official references. Its calibration groups distinguish optical structures and whether the FoV axis is specified. We evaluate its predictions by withholding each manufacturer's references in turn, testing whether the corrections also apply to devices from manufacturers absent during calibration.

The database provides the common basis for this estimation and the subsequent trend analysis. Each record identifies a hardware configuration, with specifications linked to supporting source passages and PPD estimates linked to their calculation inputs. We also use a consistent definition of head-worn mass, including the required wearing assembly and excluding components carried elsewhere. These records allow changes in display performance and hardware placement to be examined together and support updates as new configurations are documented.

Combining official PPD with SARC estimates allows us to compare resolution trends across AR glasses, VR headsets powered by a personal computer (PC), and standalone VR headsets. These categories differ in head-worn mass and FoV. To estimate annual PPD improvement at fixed mass and category, we introduce the Weight-Linked Headset Law (WL-Headset Law), which models central PPD as a function of year, head-worn mass, and device category. This provides a common comparison across successive generations: how much more central detail do newer headsets deliver at the same mass within a category? The fitted relationship also supplies a benchmark for product planning, connecting an intended launch year and mass to the resolution expected if historical progress continues.

We use this relationship to estimate when headsets could reach central PPD targets of 60 and 120 at an eyeglass-like mass of 30~g, assuming historical trends persist. The 30~g target follows the ordinary-eyeglass baseline used in a study of smart-glasses wearing comfort \cite{Kim2021SmartGlasses}. At 60 PPD, each pixel subtends one arcminute, the angular detail conventionally associated with normal visual acuity (20/20) \cite{Xiong2021Displays}. We also consider 120 PPD because measured central resolution limits reach this level in some observers \cite{Ashraf2025ResolutionLimit}. We project each category separately and combine them using an equal-category geometric mean. These trajectories describe average progress among introduced configurations; individual headsets may reach the targets earlier.

To support this analysis and quantify progress toward these targets, we make three contributions:
\begin{itemize}
    \item We compile HRMD, an open-source database containing 300 headset records, including a complete main dataset of 241 configurations from 2016--2026 used for analysis. Specifications are linked to their sources, and PPD estimates to their inputs and calculation methods, supporting verification and future updates.
    \item We develop SARC to estimate missing central PPD, reducing estimation error against official PPD from 14.8\% for traditional geometric estimation to 3.2\%. SARC provides estimates for the 175 configurations without official PPD in the main dataset.
    \item We quantify 7.4\% annual PPD growth at fixed mass and category and a 6.5\% annual decline in head-worn mass after category adjustment. At 30~g, WL-Headset Law projects average central PPD of 60 around 2035 and 120 around 2044 if historical trends persist.
\end{itemize}

\section{Related Work}\label{sec:related}
\subsection{Display and optical approaches to angular resolution}
Near-eye display research addresses the combined requirements of resolution, angular coverage, and compact optics. Reviews of AR and VR displays and light engines describe how panel technology and optical architecture jointly constrain these properties \cite{Xiong2021Displays,Hsiang2022LightEngines,Koulieris2019DisplayTracking}. Higher panel pixel densities allow more pixels without increasing display size \cite{Yin2022LiquidCrystal,Huang2020DisplayTechnologies}. Reducing the space between the panel and the eye requires additional optical changes. Maimone and Wang \cite{Maimone2020HolographicOptics} combine holographic lenses with polarization-based optical folding to demonstrate VR displays less than 9~mm thick with horizontal FoVs above 90 degrees. Bang et al.\ \cite{Bang2021LensletVR} combine a Fresnel lenslet array, a Fresnel lens, and optical folding to obtain a thin display with a 102-degree FoV in both directions. Holographic pancake optics extend this approach to optical see-through AR \cite{Cakmakci2021PancakeOptics}, while thin liquid-crystal optical components provide additional ways to implement focusing and other optical functions \cite{Xiong2021PlanarOptics,Luo2023AchromaticOptics,Chen2020GeometricPhase}.

These prototypes demonstrate concrete routes to reducing optical thickness while retaining angular coverage. Translating such progress into a lighter complete headset also depends on the supporting hardware. For example, the glasses-like prototype of Maimone and Wang places its driving electronics and light sources outside the display assembly \cite{Maimone2020HolographicOptics}. Its optical thickness and the mass of a complete head-worn system consequently describe different aspects of the design. 

Optical architecture also affects the interpretation of resolution specifications. Lens-array systems and AR combiners distribute display pixels across the view through different optical mappings \cite{Zou2022UltracompactVR,Kress2021WaveguideCombiners,Ding2023WaveguideDisplays}. PPD can therefore vary across the image, and optical aberrations can further reduce sharpness away from the center \cite{Hsiang2023MicrodisplayRequirements}. For example, Meta reports 25\% higher center sharpness and 50\% higher edge sharpness for Quest Pro with pancake optics than for Quest 2 with Fresnel optics.\footnote{\url{https://www.meta.com/blog/vr-display-optics-pancake-lenses-ppd/}} A pixel count divided by FoV does not capture all of these differences. This motivates distinguishing optical structures when estimating central PPD from published specifications, as we do in SARC.

\subsection{Visual requirements and rendering}
Display and rendering techniques address how detail is delivered. Foveated rendering concentrates computation near the gaze point and reduces peripheral rendering work \cite{Patney2016FoveatedRendering,Wang2023FoveatedSurvey}. DeepFovea uses neural reconstruction to recover images from sparse inputs \cite{Kaplanyan2019DeepFovea}, while individualized foveated rendering adapts the rendering configuration to the observer \cite{Kim2024FoveatedRendering}. These methods address the computational cost of supplying images to the display. The physical pixel count and optics still determine the angular sampling available to show those images. For a hardware forecast, distinguishing these roles connects increasing PPD to the rendering demands that future systems will need to meet.

Focus cues introduce another dimension of visual performance. Light-field displays synthesize rays corresponding to scene depths \cite{Lanman2013LightField,Huang2015LightFieldStereoscope}, while varifocal systems change the optical focus distance \cite{Dunn2017VarifocalDisplay,Aksit2017VarifocalAR}. Holographic displays form images at different depths, including through compact waveguide systems \cite{Maimone2017HolographicDisplays,Gopakumar2024HolographicAR,Jang2024WaveguideHolography}. These approaches show why a future headset may use advances in miniaturization to add visual functions as well as to reduce mass. A resolution--mass benchmark can provide a target to maintain while a manufacturer develops these capabilities.

\subsection{Head-worn mass and physical load}
Ergonomic studies show why head-worn mass is an important companion to resolution in a hardware comparison. Chihara and Seo \cite{Chihara2018PhysicalWorkload} vary headset mass and center of mass across several viewing postures to examine their effects on physical workload. Ito et al.\ \cite{Ito2021WeightBalance} study differently weighted and balanced headsets during a shooting task and find that fatigue varies with both mass and its distribution. These experiments explain why moving hardware farther away from the head or redistributing it can matter even when total mass changes little. Studies of AR interaction, neck-muscle contraction, head fixation, and neck torque further connect physical demand to task posture and headset loading \cite{Kim2020BiomechanicalStress,Zhang2023NeckMuscle,Li2025NeckMuscle,Astrologo2024NeckTorques}. Together, these findings motivate reducing mass while evaluating balance and task requirements alongside it.

\subsection{Learning from device specifications}
Estimating missing PPD also relates to regression from tabular data. Quantile regression forests (QRF) estimate conditional quantiles, including the median \cite{Meinshausen2006QuantileForests}; XGBoost learns a prediction through boosted trees \cite{Chen2016XGBoost}. ModernNCA revisits nearest-neighbor prediction with learned representations \cite{Ye2025NearestNeighbor}, and xRFM learns a feature-dependent distance metric for kernel prediction \cite{Beaglehole2026xRFM}. These approaches offer alternatives to assigning a fixed PPD ratio from pixel counts and FoV. They can use both numerical display specifications and categorical information about optical architecture. SARC uses the same available information in a more constrained form: it learns proportional corrections to geometric inputs within optical and FoV-reporting groups. This comparison asks whether the structure of the estimation problem supports accurate prediction with relatively few fitted corrections. 

\section{HMD Resolution and Mass Database}\label{sec:database}
\subsection{Scope and hardware configurations}
HRMD records display specifications and the mass of the complete head-worn assembly for each headset configuration. We collect specifications from device directories, manufacturer materials, technical reports, and reviews. The analysis uses the database snapshot dated October 13, 2026, which covers commercially available and discontinued headsets, announced devices, and research prototypes. Of its 300 records, the main analysis uses 241 complete configurations from 2016--2026, comprising 97 AR glasses, 76 PC VR headsets, and 68 standalone VR headsets from 79 manufacturers. A complete record provides the year, device category, head-worn mass, per-eye pixel counts, FoV, and optical information needed for analysis, together with official PPD or sufficient inputs to estimate it.

We choose 2016 as the starting year to align the analysis with the launch of consumer VR systems such as Oculus Rift.\footnote{\url{https://developers.meta.com/vr/blog/gdc-2025-past-present-future-developing-vr-mr-meta-audience-insights/}} Earlier coverage is much sparser: the 13 complete pre-2016 records span 1989--2015, so we retain them as historical archives and fit the trend within the more densely represented 2016--2026 period. The 46 incomplete records are retained for future completion.

We classify each configuration as AR glasses, PC VR, or standalone VR according to its display and computing architecture. AR glasses provide an optical see-through view of the surroundings. PC VR uses opaque displays and relies on an external computer for rendering. Standalone VR uses opaque displays and dedicated computing hardware, located either inside the headset or in a companion unit.

Each record represents a specific hardware version in a documented optical and wearing configuration. We merge records that differ only in product name, external controllers, or sales bundle when the head-worn hardware is identical. Distinct hardware versions, optical configurations, or head-worn assemblies remain separate records.

\subsection{Specification sources and evidence}
\begin{figure*}[!tbp]
\centering
\includegraphics[width=\textwidth]{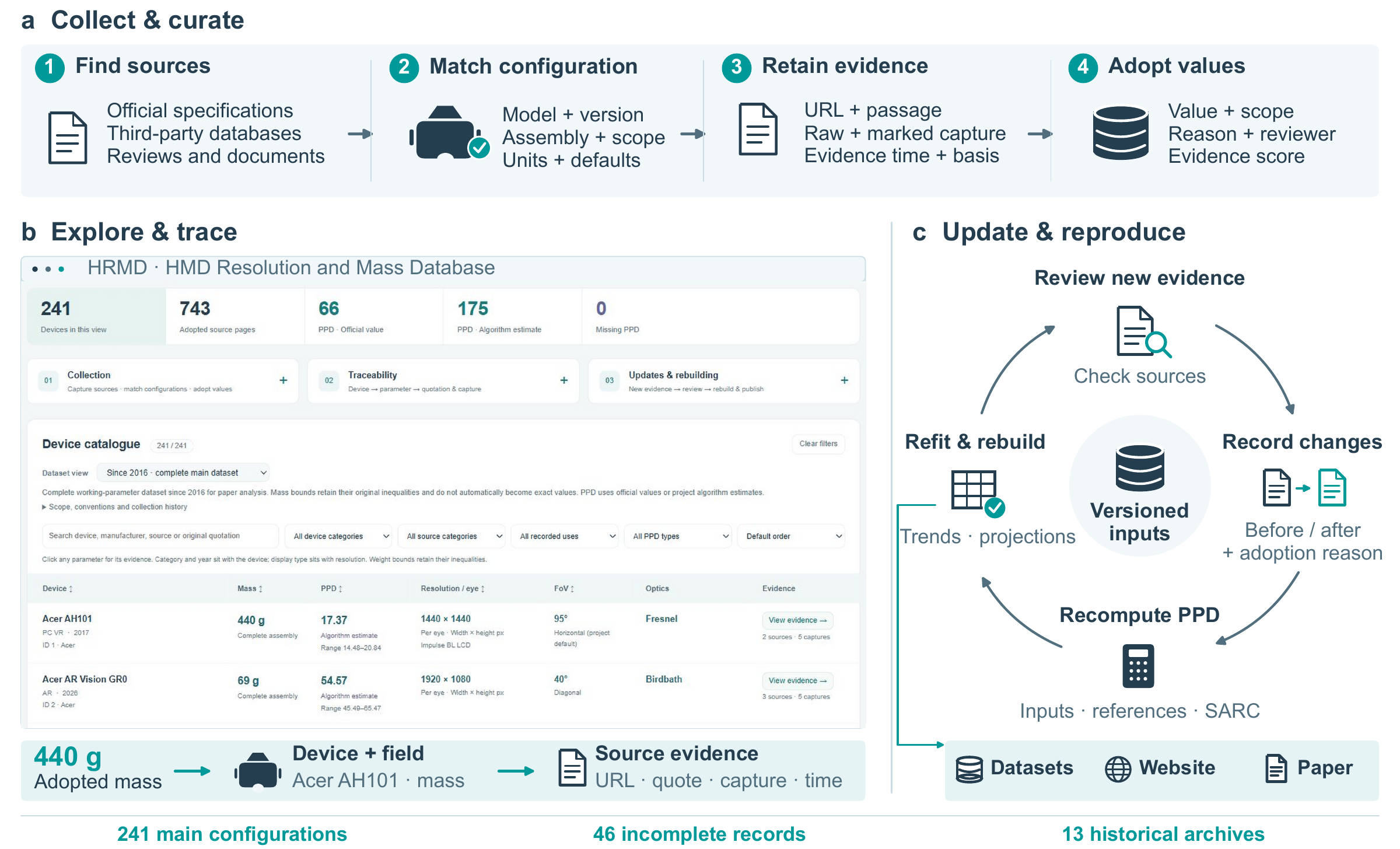}
\caption{Data collection, evidence tracing, and updating in HRMD. \textbf{(a)} Sources are collected, matched to hardware configurations, and preserved as quotations and screenshots before values are adopted with their scope and rationale. \textbf{(b)} The web catalog links device specifications to supporting evidence. \textbf{(c)} Versioned inputs are retained as new evidence is reviewed, changes are recorded, PPD is recomputed, and the trend model and projections are updated. The database contains 241 complete configurations from 2016--2026 in the main dataset, 46 incomplete records, and 13 pre-2016 historical archives.}
\label{fig:workflow}
\end{figure*}

We match each specification to the device model, hardware version, and display region or wearing configuration it describes. For pixel counts, FoV, mass, and other device specifications, matching official sources take priority, while traceable third-party sources fill gaps. Each source-reported value is linked to the supporting passage, a saved source image, and the reason for adoption. For central PPD, we adopt official values when available and use SARC estimates otherwise. The main dataset contains 66 official PPD values and 175 SARC estimates; Section~\ref{sec:accuracy} evaluates SARC's accuracy. The web catalog links specifications to their supporting evidence and PPD estimates to their inputs and calculation methods (Fig.~\ref{fig:workflow}).

\subsection{PPD reporting and analysis variables}
Only 69 of the 300 records (23.0\%) have official PPD values. Among these 69 records, the collected sources provide only a single PPD figure for 62 (89.9\%). Manufacturers may favor the typically higher central PPD because it highlights the headset's best resolution while leaving peripheral losses unquantified. This work therefore focuses on central PPD. For devices without official PPD, traditional geometric estimation divides the pixel count along an axis by the FoV along that same axis. This gives an average PPD across the view and can therefore underestimate central PPD.

The calculation also requires knowing whether the reported FoV is horizontal, vertical, or diagonal. Among the 288 records with FoV values, common reporting patterns include diagonal FoV only for 59 (20.5\%), both horizontal and vertical FoV for 42 (14.6\%), and horizontal FoV only for 19 (6.6\%). The FoV axis is unspecified for 143 records (49.7\%). SARC accounts for these reporting differences by distinguishing references with and without a specified axis within each optical group (Section~\ref{sec:estimation}).

The trend analysis uses central PPD $Q$, head-worn mass $M$ in grams, year $t$, and device category. For official PPD, we retain the stated viewing region and treat values with no stated region as central. Head-worn mass includes the headset and its required head support, facial interface, and head-mounted battery, but excludes external computing units and batteries carried elsewhere. Year records the documented event associated with the configuration, such as an announcement, public demonstration, or release. PPD estimation additionally uses horizontal and vertical pixel counts per eye, $W$ and $H$, the reported FoV $F$, and optical classification. For the FoV analysis, we convert reported angles to horizontal FoV using the stated axis. Horizontal FoV is the most frequently reported axis. We therefore treat the reported angle as horizontal by convention when its axis is unspecified. These records retain their unspecified-axis status so that SARC can calibrate them separately within each optical group.

\section{Methods}\label{sec:methods}
The analysis proceeds in three steps. We estimate missing central PPD with SARC and evaluate its accuracy on held-out official references. We then combine official and estimated PPD to fit WL-Headset Law and project resolution at specified masses. Finally, category-specific display trends and clustering help interpret differences in pixel count, angular coverage, and head-worn mass.

\subsection{Structure- and axis-aware robust calibration}\label{sec:estimation}
SARC estimates central PPD by multiplying a geometric input by a correction factor learned from official PPD. We fit separate factors for different optical structures and for records with and without a specified FoV axis.

The calibration pool contains 66 official PPD references from 21 manufacturers in HRMD's main dataset. Each reference pairs official PPD with pixel counts and FoV for the same hardware configuration and display region. For the six dual-display configurations, we use the pixel counts and FoV of the central high-density display. Their peripheral display resolution and overall FoV remain recorded separately as specifications of the complete headset.

We assign configurations to Fresnel, pancake, aspheric, small-field glasses/viewer, focal-display, and other VR groups. The Fresnel, pancake, and aspheric groups follow the documented lens architecture. The small-field group includes waveguide, birdbath, prism, and related combiner or eyepiece designs; membership follows optical structure rather than a numerical FoV cutoff. The focal-display group uses the central-region inputs described above, while the other VR group contains the remaining VR optical architectures. More detailed physical-optics classifications are also retained as inputs to the learning methods compared below.

We first calculate the traditional geometric estimate as $G=W/F_h$, where $F_h$ is horizontal FoV in degrees. For a reported horizontal FoV, we set $F_h=F$; the same convention applies when the source leaves the axis unspecified. For a rectangular display with square pixels and an explicitly reported diagonal FoV $F_d$, we calculate
\begin{equation}
F_h=2\arctan\!\left[\tan(F_d/2)\frac{W}{\sqrt{W^2+H^2}}\right].
\end{equation}
This conversion requires $0<F_d<180$ degrees; angles are converted to radians for the trigonometric calculation and the result is expressed in degrees.

We denote the geometric input used for SARC calibration by $q$. For most configurations, $q=G$. For the small-field group with a diagonal or unspecified FoV axis, we instead divide the diagonal pixel count by the reported FoV:
\begin{equation}
q_i=\begin{cases}
\sqrt{W_i^2+H_i^2}/F_i,&\substack{\text{small-field, diagonal}\\\text{or unspecified axis}};\\
G_i,&\text{otherwise}.
\end{cases}
\label{eq:sarcgeometry}
\end{equation}
These input rules apply to both calibration references and configurations requiring estimates. The FoV analysis continues to treat unspecified angles as horizontal by convention, while PPD estimation uses the calibration input $q$ defined above.

For each training reference $i$, the ratio $Q_i/q_i$ gives the correction factor needed to match its geometric input to official PPD. We calculate these corrections on the log scale as $r_i=\log_2(Q_i/q_i)$. Each reference receives weight $w_i=1/n_{g(i)}$, where $n_g$ counts all references from manufacturer $g$ in the training set $T$. This gives each manufacturer equal total weight across $T$, and these reference weights are retained when fitting individual calibration groups.

Within each optical group, we fit separate corrections for references with specified and unspecified FoV axes. We write each calibration group as $s_i=(o_i,u_i)$, where $o_i$ identifies the optical group and $u_i$ is 1 for an unspecified axis and 0 otherwise. The correction $\widehat a_s$ is the weighted median of the log ratios in group $s$, and the estimated PPD is
\begin{equation}
\begin{aligned}
\widehat a_s&\in\arg\min_a\sum_{i\in T:s_i=s}w_i|r_i-a|,\\
\widehat Q_i&=q_i2^{\widehat a_{s_i}}.
\end{aligned}
\label{eq:calibration}
\end{equation}
Thus, $2^{\widehat a_s}$ is the factor applied to geometric inputs in that group. If the weighted median is an interval rather than a single value, we use its midpoint on the log scale.

If a configuration's optical--axis group has no training references, we use the correction fitted to all references in its optical group. If that optical group also has no references, we use the correction fitted to the full training set. Both fallback corrections use the same weighted-median procedure. We fit SARC to all official references to estimate PPD for the main analysis.

\subsection{Manufacturer-held-out estimation benchmark}\label{sec:benchmark}
To test prediction accuracy for manufacturers absent from the calibration data, we hold out all references from one manufacturer at a time. In each round, we fit SARC and the learning methods using the other manufacturers, then predict PPD for the held-out manufacturer. Repeating this procedure across all 21 manufacturers gives predictions for the same 66 official PPD references under every method.

We compare SARC with traditional geometric estimation and four learning methods: QRF \cite{Meinshausen2006QuantileForests}, ModernNCA \cite{Ye2025NearestNeighbor}, XGBoost \cite{Chen2016XGBoost}, and xRFM-leaf \cite{Beaglehole2026xRFM}. QRF predicts the conditional median of PPD or its correction, depending on the selected training target. ModernNCA uses the authors' model implementation with a training procedure adapted to this dataset, and XGBoost uses the official library. Our xRFM-leaf implementation fits a single kernel model and updates its distance metric using prediction gradients.

The primary comparison uses mean absolute log$_2$ error, calculated from $|\log_2(\widehat Q_i/Q_i)|$ for each reference. An estimate twice the official PPD and an estimate half the official PPD both have an absolute log$_2$ error of 1. This gives equal penalties to overestimation and underestimation by the same multiplicative factor. We first average errors within each manufacturer and then average the manufacturer means, giving manufacturers equal weight as in SARC's calibration. A method could otherwise achieve a low overall error by performing well on manufacturers with many references, even if it performs poorly on manufacturers with few references. The mean absolute percentage error in Fig.~\ref{fig:calibration}b uses the same averaging procedure, with individual errors calculated as $100|\widehat Q_i/Q_i-1|$.

To describe accuracy across the evaluated configurations, we also report median absolute percentage error, tolerance curves, and counts within 5\%, 10\%, and 20\% error. The fraction of references whose absolute percentage error does not exceed $\tau$ is
\begin{equation}
C_m(\tau)=\frac{1}{N_R}\sum_{i=1}^{N_R}\mathbf{1}\{100|\widehat Q_{im}/Q_i-1|\leq\tau\},
\label{eq:coverage}
\end{equation}
where $N_R=66$, $m$ identifies the estimation method, and $\widehat Q_{im}$ is its prediction for reference $i$. The indicator $\mathbf{1}\{\cdot\}$ equals 1 when the prediction meets the error tolerance and 0 otherwise. Each configuration contributes equally to these summaries, so the reported proportions correspond directly to device counts.

The four learning methods receive six numerical inputs on the log$_2$ scale: $W$, $H$, $F_h$, $G$, $q$, and $W/H$. They also receive the optical group, recorded FoV axis, and detailed physical-optics classification. Training references are weighted by the inverse of their manufacturer's reference count, giving each manufacturer equal total training weight, as in SARC. Numerical standardization and categorical encoding are fitted using only the training records in each split.

Model settings are selected using only the manufacturers available for training in the current holdout round. We divide these manufacturers into three groups, train on two groups, and validate on the third, rotating through all three validation groups. Selection minimizes manufacturer-averaged absolute log$_2$ error. We consider both direct prediction of $\log_2Q$ and prediction of the log correction $r=\log_2(Q/q)$; the latter gives a PPD estimate of $q2^{\widehat r}$. SARC and its calibration variants require no hyperparameter selection.

For ModernNCA, QRF, and XGBoost, we average validation errors across multiple training runs to select model settings. We then refit the selected model multiple times and use the geometric mean of its PPD predictions. For xRFM-leaf, we fit the selected model once because repeating the fit with the same data and settings gives the same predictions.

To examine the effect of calibration grouping, we compare one correction shared by all references, separate corrections for optical groups, and separate corrections for optical--axis groups. These variants use the same geometric inputs, manufacturer weights, weighted-median calculation, and manufacturer holdouts. To test whether manufacturer weighting improves accuracy, we calculate an unweighted median of the log corrections, giving every reference equal weight. To test whether using a median improves accuracy, we retain the manufacturer weights but calculate a weighted mean of the log corrections. Both comparisons keep SARC's geometric inputs, calibration groups, fallback rules, and held-out manufacturers unchanged.

\subsection{WL-Headset Law and future projections}\label{sec:massbounds}
To quantify annual improvement in central PPD at fixed mass and device category, we fit WL-Headset Law to the 241 configurations in the main dataset. We model log PPD as a linear function of year: each additional year adds the same amount to log PPD and therefore multiplies PPD by the same factor. This describes a constant annual percentage change at fixed mass and category. We use base-two logarithms so that a doubling of PPD or mass corresponds to an increase of one on the log scale:
\begin{equation}
\begin{aligned}
\log_2 Q_i={}&a+b(t_i-2020)\\
&+c\log_2(M_i/100)+\delta_{c_i}+\epsilon_i.
\end{aligned}
\label{eq:law}
\end{equation}
Here, $i$ indexes headset configurations, $Q_i$ is central PPD, $t_i$ is year, $M_i$ is head-worn mass in grams, and $c_i$ identifies the device category. For convenience, we center year on 2020, a round year near the middle of the 2016--2026 observation period, and express mass relative to 100~g. The term $t_i-2020$ measures years relative to 2020, while $\log_2(M_i/100)$ measures mass in doublings relative to 100~g, taking values of $-1$, $0$, and $1$ for masses of 50, 100, and 200~g, respectively. Both terms are zero for a 100~g headset in 2020. The intercept $a$ therefore gives the fitted log$_2$ PPD for the reference category at that year and mass, while $\delta_{c_i}$ gives the difference for category $c_i$ relative to the reference category and $\epsilon_i$ is the residual. Using a different reference year or mass would change the intercept but leave the estimated annual growth rate and fitted PPD unchanged.

We fit Eq.~\eqref{eq:law} by ordinary least squares to the 241 configurations in the main dataset, giving each configuration equal weight and using AR glasses as the reference category. At fixed mass and category, the fitted year coefficient $b$ gives an annual PPD change of $100(2^b-1)\%$. At fixed year and category, doubling mass multiplies fitted PPD by $2^c$. The model predicts log$_2$ PPD, which we convert back to PPD; for example, a predicted log value of 5 corresponds to 32 PPD. The resulting prediction represents the geometric-mean PPD for headsets of a given year, mass, and category.

To compare historical growth rates across categories, we also fit a model with a separate year coefficient for each category while retaining a common mass coefficient. We convert each year coefficient to an annual percentage change before calculating differences between categories in percentage points. A separate regression of $\log_2(M/100)$ on year and category estimates the annual change in head-worn mass after accounting for category differences.

\subsubsection{Future projections}
Future projections use the primary model in Eq.~\eqref{eq:law}, with its fitted coefficients held constant beyond 2026. Each category retains its fitted offset, while all three share the same annual growth rate and mass coefficient. For the overall projection, we use the average of the three category offsets, denoted by $\overline{\delta}$. This gives the geometric mean of the three category predictions, with equal category weights regardless of their numbers of configurations in the dataset.

For a target central PPD $Q_*$ at mass $M_*$, we rearrange Eq.~\eqref{eq:law} to calculate when the equal-category geometric mean reaches that target:
\begin{equation}
t_*=2020+\frac{\log_2Q_*-\widehat a-\widehat c\log_2(M_*/100)-\overline{\delta}}{\widehat b}.
\label{eq:projection}
\end{equation}
Replacing $\overline{\delta}$ with the fitted offset for an individual category gives that category's target year. We evaluate this expression at 30~g for targets of 60 and 120 PPD. Varying $M_*$ gives the mass--year contours in Fig.~\ref{fig:atlas}a, while fixing mass at 30~g and varying year gives the PPD trajectories in Fig.~\ref{fig:atlas}b.

\subsubsection{Uncertainty in the projections}
To calculate uncertainty around these trajectories, we generate 2,000 bootstrap samples by sampling manufacturers with replacement and retaining all configurations belonging to each sampled manufacturer. This keeps configurations from the same manufacturer together. In each sample, we refit SARC using the resampled official references, update the estimated PPD values, and refit WL-Headset Law.

We also account for the remaining error in SARC estimates by adding independent, zero-mean normal noise to their log$_2$ PPD values. Its standard deviation is $\sigma_Q=0.134$, obtained by averaging squared held-out log errors within each manufacturer and then across manufacturers before taking the square root. Refitting SARC allows estimates that share a calibration factor to change together, while the added noise represents variation remaining after calibration. Official PPD values remain fixed.

The main analysis uses the mass values adopted in HRMD. Where only an upper bound is reported, we use that bound as a mass approximation. During bootstrap resampling, masses with finite reported ranges are drawn uniformly within those ranges; all other mass values remain fixed.

Each valid bootstrap refit supplies a PPD trajectory for every category and their equal-category geometric mean. The mean trajectory uses the average category offset from that same refit. At each year, the 2.5th and 97.5th percentiles of the predicted PPD values define the pointwise 95\% confidence bands in Fig.~\ref{fig:atlas}b.

\subsection{Annual display trends and design clusters}\label{sec:annual}
For the annual trends in Fig.~\ref{fig:pixelsfield}, we separately regress the natural logarithms of central PPD, horizontal FoV, and horizontal pixels per eye on year within each device category. We use ordinary least squares with equal configuration weights and no mass adjustment. A fitted year coefficient $\beta$ corresponds to an annual percentage change of $100(e^\beta-1)\%$.

To examine how accounting for head-worn mass changes the estimated FoV trend, we fit two models to the same configurations. Both regress $\log_2F_h$ on year and category, but the second also includes $\log_2(M/100)$. Comparing their year coefficients gives the difference between the category-adjusted annual trend and the trend at fixed mass and category.

To identify groups of configurations with similar mass and PPD in Fig.~\ref{fig:landscape}, we apply $k$-means clustering to $\log_2M$ and $\log_2Q$. Before clustering, we subtract each log variable's mean and divide by its population standard deviation, giving the two variables the same scale. For each candidate $k=2,\ldots,8$, we use Lloyd's algorithm with 100 initializations, at most 1,000 iterations, and a convergence tolerance of $10^{-8}$. We select the number of clusters using the mean silhouette score, which favors configurations that are close together within each cluster and well separated from other clusters. The highest score is obtained with three clusters. We label them P1--P3 in order of increasing geometric-mean mass.

For the mass--FoV clustering in Fig.~\ref{fig:designclusters}, we standardize $\log_2M$ and $\log_2F_h$ by subtracting each variable's mean and dividing by its population standard deviation. We apply $k$-means with $k=5$, using the same initialization, iteration, and convergence settings as for the mass--PPD clustering described above. Clusters are labeled C1--C5 in order of increasing geometric-mean mass.

\section{Results}\label{sec:results}
\subsection{Resolution and mass across headset configurations}\label{sec:landscape-results}
\begin{figure*}[!tbp]
\centering
\includegraphics[width=\textwidth]{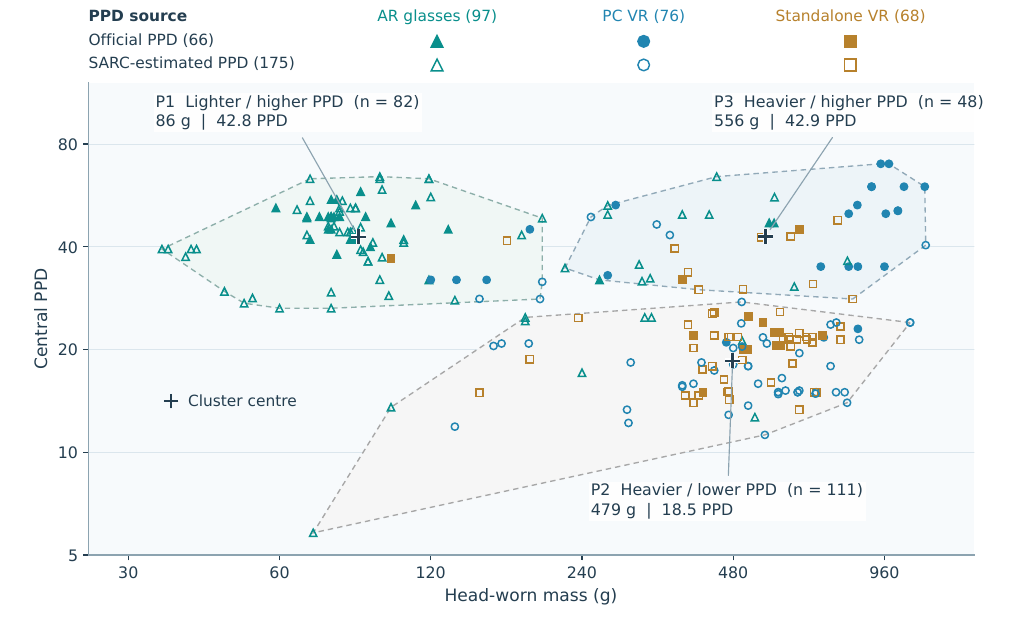}
\caption{Central PPD and head-worn mass of the 241 configurations in HRMD's main dataset, plotted on logarithmic axes. Marker color and shape distinguish AR glasses, PC VR headsets, and standalone VR headsets. Filled and open markers indicate official PPD (66 configurations) and SARC-estimated PPD (175 configurations), respectively. P1--P3 are clusters obtained by applying $k$-means to standardized log mass and log PPD (Section~\ref{sec:annual}). Dashed outlines enclose the configurations in each cluster, and plus signs mark their geometric-mean mass and PPD.}
\label{fig:landscape}
\end{figure*}

The three mass--PPD clusters in Fig.~\ref{fig:landscape} highlight substantial differences in central PPD among headsets of similar mass. For P2, the geometric means are 478.7~g and 18.5 PPD; for P3, they are 556.4~g and 42.9 PPD. The lightest cluster, P1, consists mainly of AR glasses and has geometric means of 86.1~g and 42.8 PPD, nearly matching P3 in central PPD despite its much lower mass. These differences motivate accounting for both head-worn mass and device category when estimating PPD improvement over time. Because this comparison relies on estimated PPD for 175 configurations, we next assess how accurately SARC predicts the official references.

\subsection{Accuracy of central PPD estimation}\label{sec:accuracy}
\begin{figure*}[!tbp]
\centering
\includegraphics[width=\textwidth]{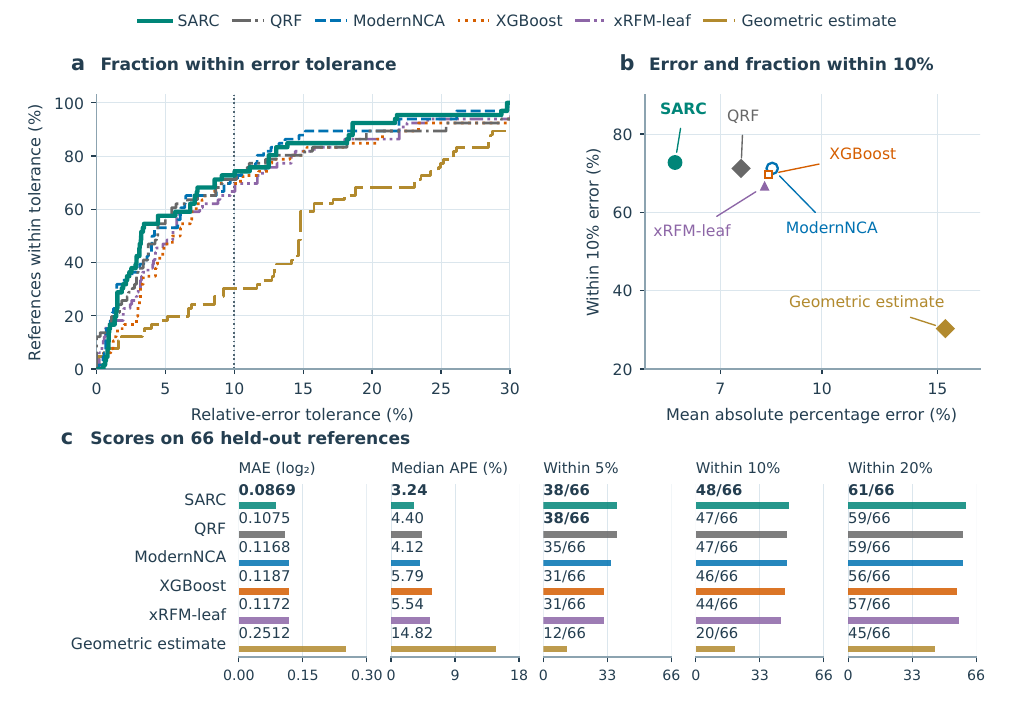}
\caption{Accuracy of central PPD estimates against 66 official references held out by manufacturer. \textbf{(a)} Percentage of references within each absolute relative-error tolerance from 0 to 30\%; higher curves indicate greater accuracy, and the dotted line marks the 10\% tolerance. \textbf{(b)} Mean absolute percentage error (horizontal axis, logarithmic scale) and percentage of references within 10\% error (vertical axis); better performance lies toward the upper left. \textbf{(c)} Mean absolute log$_2$ error (MAE), median absolute percentage error (APE), and reference counts within 5\%, 10\%, and 20\% error. Bold values indicate the best result for each metric, including ties.}
\label{fig:calibration}
\end{figure*}

We evaluate SARC by predicting each manufacturer's official PPD values using only references from other manufacturers. Compared with traditional geometric estimation, SARC reduces median absolute percentage error from 14.8\% to 3.2\% and increases the proportion of estimates within 10\% error from 30.3\% to 72.7\% (Fig.~\ref{fig:calibration}). SARC estimates 48 of the 66 configurations within 10\% of official PPD. Fitting SARC to all official references supplies the estimates used in the subsequent trend analysis.

Beyond traditional geometric estimation, we compare SARC with the four learning methods introduced in Section~\ref{sec:benchmark}, using the same display and optical information. SARC achieves the lowest mean absolute log$_2$ error at 0.087, compared with 0.107 for QRF, 0.117 for ModernNCA and xRFM-leaf, and 0.119 for XGBoost (Fig.~\ref{fig:calibration}c). This is a 19.1\% reduction relative to QRF, the best-performing external baseline on this metric. SARC achieves this result without hyperparameter tuning, estimating a single multiplicative correction within each optical and FoV-reporting group.

To understand which parts of SARC contribute to this accuracy, we compare alternative ways of grouping references and calculating their corrections. Applying one correction to all optical structures assumes that their geometric estimates require the same proportional adjustment. This gives a mean absolute log$_2$ error of 0.196; allowing each optical group to have its own correction reduces the error to 0.104, and further separating references by whether the FoV axis is specified reduces it to 0.087. With geometry, groups, and validation folds unchanged, replacing SARC's weighted median with a manufacturer-weighted mean raises mean absolute log$_2$ error from 0.087 to 0.107. SARC applies the same correction factor to all devices in a group. If a few references have official PPD much higher than their geometric estimates, using a mean can increase this factor and push up the estimates for other devices in the group as well. A median is less affected by these extreme references, reducing their influence on estimates for the rest of the group. Separately, retaining the median but removing manufacturer weights raises error from 0.087 to 0.096. Using official PPD and SARC estimates, we next examine how headset resolution and head-worn mass have changed over the decade.

\subsection{A decade of weight-linked progress}\label{sec:progress}
Across 2016--2026, the fitted year coefficient $b$ corresponds to a 7.4\% annual increase in central PPD at fixed mass and category. For headsets of the same category and year, the model associates a doubling of head-worn mass with 5.9\% higher central PPD. We also examine how head-worn mass changes over time by fitting mass against year and device category. This separate model estimates a 6.45\% annual decline in head-worn mass after accounting for category differences.

To examine how the PPD trend differs across categories, we allow each category to have its own annual growth rate. At fixed mass, the estimated annual increases are 10.7\% for PC VR, 9.0\% for standalone VR, and 2.0\% for AR glasses. The PC VR growth rate exceeds that of AR glasses by 8.7 percentage points per year.

\subsection{Pixel growth, field of view, and design variation}\label{sec:design-results}
\begin{figure*}[!tbp]
\centering
\includegraphics[width=\textwidth]{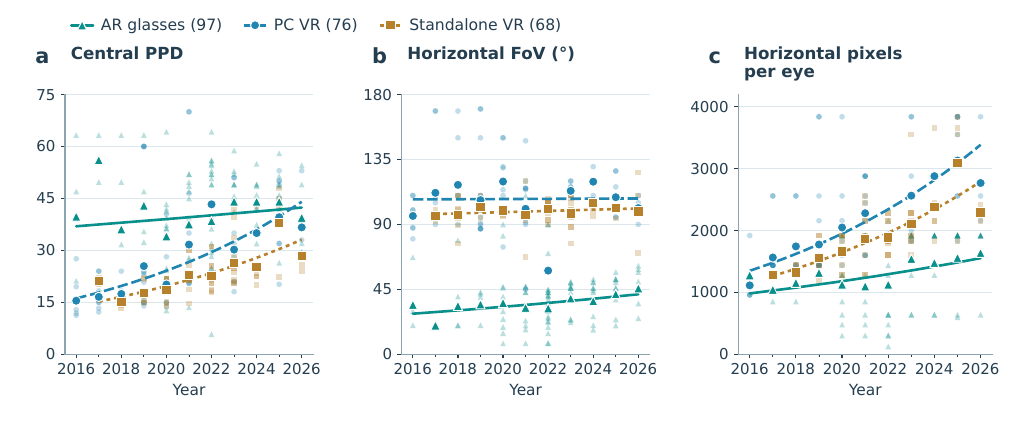}
\caption{Annual trends in central PPD, horizontal FoV, and horizontal pixel count for the 241 configurations in HRMD's main dataset. \textbf{(a)} Central PPD, using official values or SARC estimates. \textbf{(b)} Horizontal FoV. \textbf{(c)} Horizontal pixels per eye. Small markers represent individual configurations; large markers show the geometric mean for each category and year. Curves show log-linear trends fitted separately for each category, giving each configuration equal weight without adjusting for head-worn mass (Section~\ref{sec:annual}). Each curve covers the years observed for its category. Colors, marker shapes, and line styles distinguish device categories.}
\label{fig:pixelsfield}
\end{figure*}

To understand why central PPD grows faster in VR than in AR glasses, we compare changes in horizontal pixels per eye and horizontal FoV alongside central PPD (Fig.~\ref{fig:pixelsfield}). In PC VR and standalone VR, horizontal pixel counts increase by 9.6\% and 9.2\% per year, respectively, while horizontal FoV increases by less than 0.5\% per year in both categories. VR headsets therefore provide more pixels within a nearly unchanged angular span. In AR glasses, horizontal pixel count increases by 4.7\% per year, accompanied by a 4.0\% annual increase in horizontal FoV. This field expansion spreads the pixels over a wider angle, helping explain why pixel growth produces a smaller increase in PPD.

\begin{figure*}[!tbp]
\centering
\includegraphics[width=\textwidth]{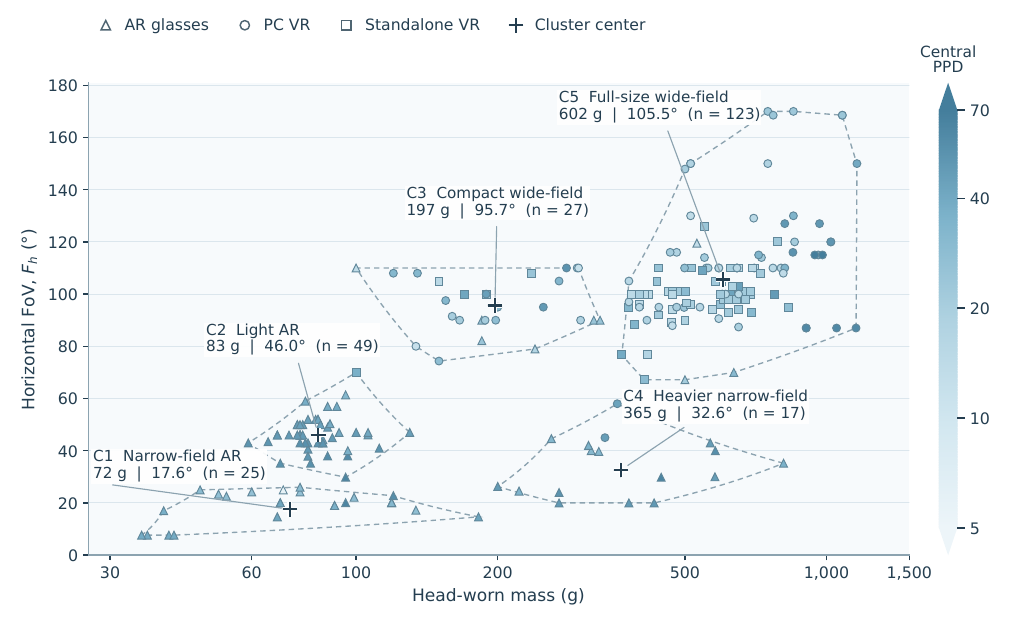}
\caption{Head-worn mass, horizontal FoV, and central PPD for the 241 configurations in HRMD's main dataset. Marker position shows mass on a logarithmic axis and horizontal FoV on a linear axis. Color indicates official or SARC-estimated central PPD, and shape identifies device category. Dashed outlines enclose five clusters, C1--C5, obtained by applying $k$-means to standardized log mass and log FoV (Section~\ref{sec:annual}). Plus signs mark cluster centers, given by the geometric means of mass and FoV; labels show these values and the number of configurations.}
\label{fig:designclusters}
\end{figure*}

The category-specific trends above include changes in the mass of the headsets introduced each year. Across all three categories, the estimated annual increase in horizontal FoV is 1.56\% after accounting for category differences. Holding mass fixed as well raises this estimate to 3.14\%. Within the same year and category, lighter headsets tend to have narrower FoVs.

Figure~\ref{fig:designclusters} shows the different combinations of mass and angular coverage underlying these trends. The two light AR clusters, C1 and C2, are centered at 72~g and 17.6$^\circ$, and 83~g and 46.0$^\circ$, respectively: their masses are similar, but their FoVs differ substantially. The compact and full-size wide-field clusters, C3 and C5, have similar FoV centers of 95.7$^\circ$ and 105.5$^\circ$, but mass centers of 197 and 602~g. High central PPD occurs in both light AR glasses with narrow FoVs and heavier headsets with wide FoVs.

\subsection{Projecting future headset evolution}\label{sec:projection-results}
\begin{figure*}[!tbp]
\centering
\includegraphics[width=\textwidth]{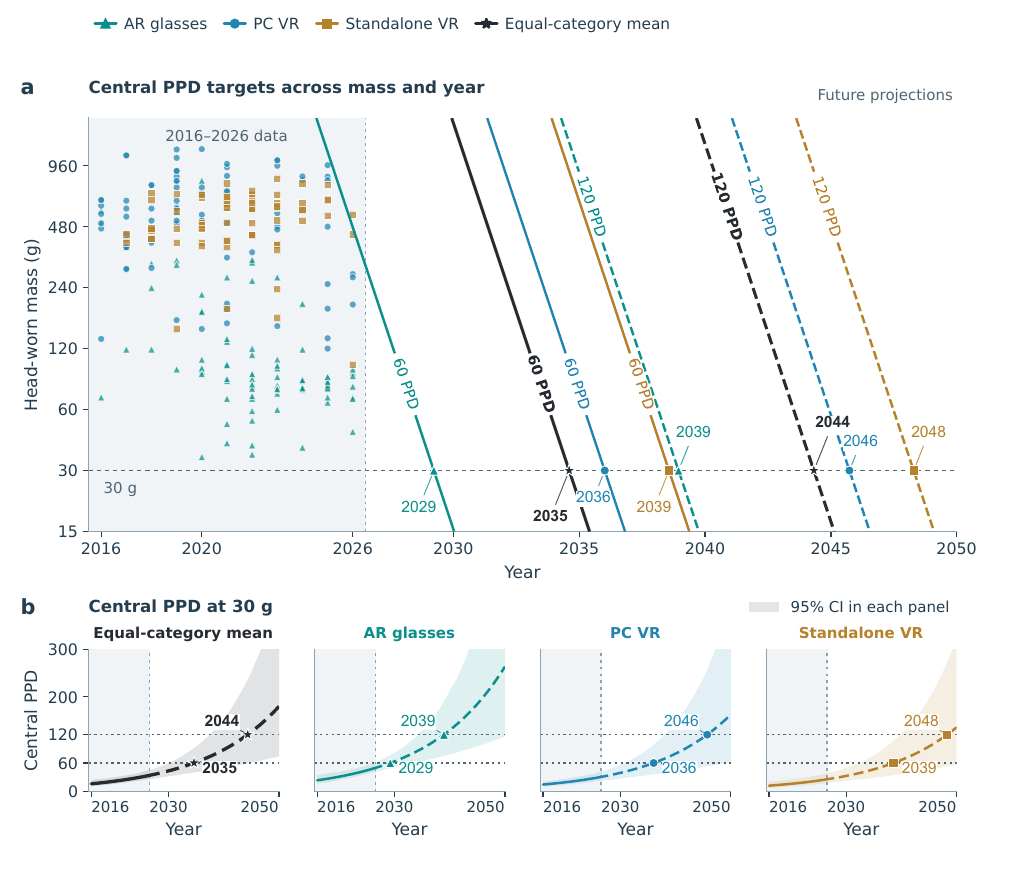}
\caption{Central PPD projections from WL-Headset Law for AR glasses, PC VR, standalone VR, and their equal-category geometric mean. \textbf{(a)} Solid and dashed contours show mass--year combinations reaching 60 and 120 PPD, respectively, with mass on a logarithmic axis. Colored contours show individual categories; black contours show their mean. Small markers show the year and head-worn mass of observed configurations, with color and shape identifying category. The horizontal dotted line marks 30~g. \textbf{(b)} Central PPD at 30~g for the mean and each category, shown on shared linear scales. Solid segments cover the observation period, and dashed segments project beyond it. Shading around each curve shows its pointwise 95\% manufacturer-bootstrap confidence interval; horizontal dotted lines mark 60 and 120 PPD. In both panels, symbols mark the target crossings at 30~g, with stars for the mean and year labels connected by thin lines. The pale background marks 2016--2026, separated from future projections by a vertical dashed line. All projections use the same fitted time and mass coefficients, with category-specific offsets.}
\label{fig:atlas}
\end{figure*}

To estimate when headsets could reach 60 and 120 central PPD at 30~g, we extrapolate WL-Headset Law beyond the 2016--2026 observation period, assuming historical trends persist. Substituting the fitted coefficients gives
\begin{equation*}
\begin{aligned}
\log_2\widehat Q={}&5.10283+0.10284(t-2020)\\
&+0.08304\log_2(M/100)+\widehat\delta_c,
\end{aligned}
\end{equation*}
where $\widehat Q$ is predicted central PPD and $\widehat\delta_c$ is 0 for AR glasses, $-0.69843$ for PC VR, and $-0.96102$ for standalone VR. We use these offsets to project each category separately and calculate the geometric mean of the three predictions with equal category weights. Figure~\ref{fig:atlas}a shows the combinations of mass and year at which each category and their average reach 60 and 120 PPD.

At 30~g, the projected average central PPD reaches 60 around 2035 and 120 around 2044 (Fig.~\ref{fig:atlas}b). AR glasses reach these targets earliest, around 2029 and 2039, respectively. PC VR follows around 2036 and 2046, and standalone VR around 2039 and 2048. The shaded bands widen toward later years because uncertainty in the estimated annual growth rate has a larger cumulative effect over longer projections.

For headsets introduced in the same year and with the same head-worn mass, the model estimates that AR glasses have 62\% higher central PPD than PC VR and 95\% higher central PPD than standalone VR. We assume that all three categories increase their PPD by 7.4\% per year in these projections. At this shared growth rate, the higher PPD predicted for AR glasses means that they reach 60 and 120 PPD first, followed by PC VR and standalone VR.

The earlier AR crossings are consistent with the high central PPD and narrow FoVs of the light AR configurations in Fig.~\ref{fig:designclusters}. Concentrating pixels within a smaller angular field supports high PPD, whereas VR headsets distribute pixels across a wider view. The later standalone trajectory may also reflect how head-worn mass is allocated: many standalone headsets carry computing and battery hardware on the head, whereas PC VR offloads rendering to a separate computer.

\section{Discussion}\label{sec:discussion}
\subsection{Average trajectories and early products}
WL-Headset Law provides a reference for the average central PPD of headsets introduced in a given year, at a specified mass and within each device category. Our overall projection combines the three category predictions using an equal-category geometric mean. The projected crossings around 2035 and 2044 therefore describe when this average reaches 60 and 120 PPD at 30~g; individual products may achieve these combinations years earlier. Varjo illustrates how selected products can reach a resolution milestone well before it becomes typical: the VR-1 already offered more than 60 PPD in its central region when introduced in 2019.\footnote{\url{https://varjo.com/news/varjo-announces-vr-1-worlds-first-human-eye-resolution-vr-headset-for-industrial-use}} The later VR-3 provides 70 PPD within a central region of $27\times27$ degrees.\footnote{\url{https://varjo.com/products/varjo-vr-3}} For such heavier, high-resolution headsets, the route toward our joint target involves retaining that detail while reducing head-worn mass. The distinction between an early product and the average trajectory is therefore important when interpreting the projected dates.

\subsection{Using the law in product planning}
Manufacturers can use this trajectory to set resolution goals for a planned release. Given the intended launch year, head-worn mass, and category, WL-Headset Law supplies a projected average PPD against which the proposed specification can be compared. At the fitted annual growth rate of 7.4\%, this benchmark rises by about 24\% over three years at the same mass and category. A product designed to match today's average PPD could therefore fall below the projected average by the end of a three-year development cycle. Using the planned launch date helps manufacturers set a target that accounts for improvement elsewhere in the industry. Repeating this comparison across a product roadmap can show whether successive generations are expected to keep pace with the historical trend or establish a larger lead. Supporting these resolution targets also requires planning the rendering workload. Foveated rendering provides one approach by concentrating rendering detail near the gaze point and reducing peripheral rendering work \cite{Patney2016FoveatedRendering,Wang2023FoveatedSurvey,Kaplanyan2019DeepFovea,Kim2024FoveatedRendering}.

For a product already projected to exceed its category's PPD benchmark, a manufacturer could set a further goal of adding visual capabilities while preserving that resolution and head-worn mass. Light-field and varifocal displays offer ways to provide depth-dependent focus cues \cite{Lanman2013LightField,Huang2015LightFieldStereoscope,Dunn2017VarifocalDisplay,Aksit2017VarifocalAR}, while holographic displays form images at multiple depths \cite{Maimone2017HolographicDisplays}. Holographic waveguide prototypes demonstrate approaches to incorporating this capability into compact AR optics \cite{Gopakumar2024HolographicAR,Jang2024WaveguideHolography}. A manufacturer considering these architectures can use the launch-year PPD benchmark as a resolution requirement during development, alongside the desired focus behavior and mass target. This gives the law a practical role in evaluating a new architecture: whether it can add the intended capability while retaining the resolution expected of its future category.

As new headsets are introduced, these comparisons can also distinguish isolated advances from changes across the industry. A product above the fitted trajectory demonstrates a combination of mass and PPD that exceeds the historical average. If later generations from multiple manufacturers repeatedly exceed the projected values, the updated data may support a faster rate of progress or a change in the differences between categories. HRMD allows the fitted relationship and future benchmarks to be revised as those specifications accumulate. Manufacturers can consequently use WL-Headset Law throughout development: to set an initial target, reassess that target as competing devices appear, and evaluate how their released products compare with the evolving category average.

\subsection{Different priorities across device categories}
The category results help manufacturers interpret these comparisons and identify opportunities for improvement. AR glasses reach the projected targets earlier because of their higher fitted PPD at a given mass, even though their historical PPD growth has been slower than that of VR. Their expanding FoV shows that display improvements have also increased the area available for virtual content. An AR manufacturer could therefore use the PPD benchmark as a level to maintain while developing a wider field, rather than making higher central PPD the sole goal of each generation. PC and standalone VR have increased PPD more rapidly while retaining a nearly constant FoV. For these categories, reducing head-worn mass while preserving that coverage offers another direction for product differentiation. Standalone VR also carries computing and energy-storage requirements that can be addressed through changes in hardware placement. These category differences turn the common PPD--mass target into different development priorities.

Meta VR Glasses provides a concrete example of such an architectural change. The announced device combines approximately 100~g of head-worn hardware with 37 central PPD, while placing computing, storage, and the battery in a separate puck.\footnote{\url{https://www.meta.com/blog/meta-vr-glasses-announcement-meta-connect/}} PICO's separate headset and main-unit regulatory filings suggest a similar division of hardware; its announced Space Pro display targets average PPD of 40 and central PPD above 45.\footnote{\url{https://developer.picoxr.com/document/discover/xr-devices/}} For Meta VR Glasses, using its 2026 announcement year and 100~g head-worn mass, WL-Headset Law gives a fitted standalone-VR average of approximately 27 PPD. The announced 37 PPD is about 37\% above that benchmark. This comparison shows how the law can quantify a product's position relative to the historical average. Its 70-degree horizontal FoV is also the fourth narrowest among all 144 VR configurations in HRMD's main dataset.\footnote{\url{https://developers.meta.com/vr/essentials/field-of-view/}} The design thus combines performance above the fitted PPD benchmark with off-head computing and relatively narrow angular coverage. A manufacturer pursuing similar mass and PPD with a wider FoV would have a further requirement to address through its display and optical design.

\section{Limitations and Future Work}\label{sec:limitations}
HRMD draws on published specifications with different reporting conventions for PPD, FoV, and wearing configurations. Many PPD values do not identify the region of the view they describe, and many FoV values omit the axis. The main analysis requires enough published information to determine both central PPD and head-worn mass. Devices with incomplete specifications are therefore absent from the fitted trends. Comparable sales and usage figures are not routinely disclosed for individual headset models, so the analysis gives each configuration equal weight. Products represented by multiple configurations therefore contribute more to the fit.

The projections apply the same annual PPD growth rate of 7.4\% to all three categories and assume that the relationship with mass and the differences between categories remain unchanged. However, the historical results show faster PPD growth in VR than in AR. If this difference continues, the PPD gaps between categories will change, affecting the predicted target years. The shaded intervals in Fig.~\ref{fig:atlas} reflect variation across manufacturers and uncertainty in SARC estimates and recorded mass. They do not include future changes in display technology, hardware architecture, or product priorities that could alter the rate of progress.

As evidence accumulates, the database can support re-estimation of category-specific trajectories and a more detailed account of where high resolution is provided across the view. More consistent reporting of central and peripheral PPD, FoV axes, and wearing configurations would make those comparisons easier. Repeating the benchmark on newly documented manufacturers would also test whether the current optical-group corrections continue to apply.

\section{Conclusion}
Combining official PPD with SARC estimates in HRMD allows us to trace a decade of resolution and head-worn mass changes. WL-Headset Law estimates a 7.4\% annual increase in central PPD at fixed mass and category. Over the same period, head-worn mass declined by 6.5\% annually after accounting for category differences. If historical trends persist, the equal-category geometric mean of central PPD at an eyeglass-like mass of 30~g is projected to reach 60 around 2035 and 120 around 2044. These PPD targets represent central sampling levels associated with normal visual acuity (20/20) and higher visual acuity, respectively. The resulting benchmarks can help manufacturers set resolution targets for a planned launch year, mass, and category, and can be revised as HRMD incorporates new devices.

\section*{Data and Code Availability}
HRMD device records and field-level source evidence, together with the SARC implementation, calculation inputs, and documentation of the search and evidence-collection process, are publicly available at \url{https://github.com/chaosikaros/hmd-resolution-mass-database}.

\bibliographystyle{abbrv-doi}
\bibliography{selected_references}
\end{document}